\documentclass[aps,prb,superscriptaddress,showpacs,reprint]{revtex4-1}
\pdfoutput=1
\usepackage{color}
\usepackage{natbib}
\usepackage{amsmath}
\usepackage{amssymb}
\usepackage[pdftex]{graphicx}
\usepackage[ngerman, english]{babel} 
\usepackage{siunitx}
\usepackage[version =3]{mhchem}
\usepackage{soul}	
\usepackage{color}
\usepackage{lineno}
\usepackage[T1]{fontenc}

\usepackage{lipsum}
\usepackage{hyperref}

\usepackage{xr}
\newcommand{\EF}{$E_{\rm F}$\;}

\newcommand{\Aa}{$a_{\rm 0}$\;}
\newcommand{\EB}{$E-E_{\rm F}$\;}

\newcommand{\hv}{$h\nu$\;}
\newcommand{\VTM}{$V_{\rm Tm}$\;}
\newcommand{\TM}{TmSe$_{\rm 1-x}$Te$_{\rm x}$\;}
\newcommand{\nf}{$n_{\rm f}$\;}

\newcommand{\HS}{$^{\rm 3}H_{\rm 6}$\;}
\newcommand{\Ff}{$^{\rm 3}F_{\rm 4}$\;}

\begin{document}
%\title{Coherent many-body resonance state forming indirect band gaps  in mixed-valence \TM}
%\title{Two exotic many-body resonance peaks and conduction band deformation in mixed-valence \TM}
%\title{Near-gap spectra and Tm valence of mixed-valence \TM}
%\title{}
\title{Anomalous 4$f$ fine structure in \TM across the metal-insulator transition}

\author{C.-H. Min}
\email[corresponding author for photoemission analyses. Email:]{chul.h.min@ntnu.no}
\affiliation{Institut f\"ur Experimentelle und Angewandte Physik, Christian-Albrechts-Universit\"at zu Kiel and Kiel Nano, Surface and Interface Science KiNSIS, D--24098 Kiel, Germany}
\affiliation{Center for Quantum Spintronics, Department of Physics, Norwegian University of Science and Technology, NO7034 Trondheim, Norway}

\author{S. M\"uller}
\affiliation{Experimentelle Physik VII and W\"urzburg-Dresden Cluster of Excellence ct.qmat, Fakult\"at f\"ur Physik und Astronomie, Universit\"at W\"urzburg,  D--97074 W\"urzburg, Germany}

\author{W. J. Choi}
\affiliation{Department of Physics and Chemistry, DGIST, Daegu 42988, Republic of Korea}

\author{L. Dudy}
\affiliation{Synchrotron SOLEIL,  Saint-Aubin 91190, France}

\author{V. Zabolotnyy}
\affiliation{Experimentelle Physik IV, Universit\"at W\"urzburg,  D--97074 W\"urzburg, Germany}

\author{M. Heber}
\affiliation{Deutsches Elektronen-Synchrotron DESY, D--22607 Hamburg, Germany}

\author{J. D. Denlinger}
\affiliation{Advanced Light Source, Lawrence Berkeley Laboratory, Berkeley, California 94720, USA}

\author{C.-J. Kang}
\affiliation{Department of Physics, Chungnam National University, Daejeon 34134, Republic of Korea}

\author{M. Kall\"ane}
\affiliation{Institut f\"ur Experimentelle und Angewandte Physik, Christian-Albrechts-Universit\"at zu Kiel and Kiel Nano, Surface and Interface Science KiNSIS, D--24098 Kiel, Germany}
\affiliation{Ruprecht Haensel Laboratory, Christian-Albrechts-Universit\"at zu Kiel, D--24098 Kiel, Germany }

\author{N. Wind}
\affiliation{Institut f\"ur Experimentelle und Angewandte Physik, Christian-Albrechts-Universit\"at zu Kiel and Kiel Nano, Surface and Interface Science KiNSIS, D--24098 Kiel, Germany}
\affiliation{Deutsches Elektronen-Synchrotron DESY, D--22607 Hamburg, Germany}
\affiliation{Institut f\"ur Experimentalphysik, Universit\"at Hamburg, 22761 Hamburg, Germany}

\author{M. Scholz}
\affiliation{Deutsches Elektronen-Synchrotron DESY, D--22607 Hamburg, Germany}

\author{T. L. Lee}
\affiliation{Diamond Light Source, Harwell Science and Innovation Campus, Didcot, OX11 0DE, UK}

\author{C. Schlueter} 
\affiliation{Deutsches Elektronen-Synchrotron DESY, D--22607 Hamburg, Germany}

\author{A. Gloskovskii}
\affiliation{Deutsches Elektronen-Synchrotron DESY, D--22607 Hamburg, Germany}

\author{E. D. L. Rienks}
\affiliation{Helmholtz-Zentrum Berlin f‌\"ur Materialien und Energie, Elektronenspeicherring BESSY II, D--12489 Berlin, Germany.}

\author{V. Hinkov}
\affiliation{Experimentelle Physik IV, Universit\"at W\"urzburg, D--97074 W\"urzburg, Germany}

\author{H. Bentmann}
\affiliation{Center for Quantum Spintronics, Department of Physics, Norwegian University of Science and Technology, NO7034 Trondheim, Norway}

\author{Y. S. Kwon}
\email[corresponding author for the single-crystal growth and characterization. Email:]{yskwon@dgist.ac.kr}
\affiliation{Department of Physics and Chemistry, DGIST, Daegu 42988, Republic of Korea}

\author{F. Reinert}
\affiliation{Experimentelle Physik VII and W\"urzburg-Dresden Cluster of Excellence ct.qmat, Fakult\"at f\"ur Physik und Astronomie, Universit\"at W\"urzburg,  D--97074 W\"urzburg, Germany}

\author{K. Rossnagel}
\affiliation{Institut f\"ur Experimentelle und Angewandte Physik, Christian-Albrechts-Universit\"at zu Kiel and Kiel Nano, Surface and Interface Science KiNSIS, D--24098 Kiel, Germany}
\affiliation{Ruprecht Haensel Laboratory, Christian-Albrechts-Universit\"at zu Kiel, D--24098 Kiel, Germany }
\affiliation{Ruprecht Haensel Laboratory, Deutsches Elektronen-Synchrotron DESY, D--22607 Hamburg, Germany }
%\date{\today}

%% key word: Tm valence, gap opening, coherent many-body state, mixed valence, indirect band gap, ARPES, topology, excitonic state, large Fermi surface, coherent 4f-hole band,    

\begin{abstract}

Hybridization between localized 4$f$ and itinerant 5$d$6$s$ states in heavy fermion compounds is a well-studied phenomenon and commonly captured by the paradigmatic Anderson model. However, the investigation of additional electronic interactions, beyond the standard Anderson model, has been limited, despite their predicted important role in the exotic quasiparticle formation in mixed-valence systems. We investigate the 4$f$ states in \TM throughout a semimetal-insulator phase transition, which drastically varies the interactions related to the 4$f$ states. Using synchrotron-based hard x-ray and extreme ultraviolet photoemission spectroscopy, we resolve subtle peak splitting in the 4$f$ peaks near the Fermi level in the mixed-valent semimetal phase. The separation is enhanced by several tens of meV by increasing the lattice parameter by a few percent. Our results elucidate the evolving nature of the 4$f$ state across the phase transition, and provide direct experimental evidence for electronic interactions beyond the standard Anderson model in mixed-valence systems.

\end{abstract} 
\maketitle 
%\section{Introduction}

Elucidating the mechanism of quasiparticle formation is essential for gaining an intuitive understanding of complex electronic states in many-body systems \cite{landau1959theory}. Based on the Landau-Fermi liquid theory, complex many-electron problems are simplified by representing them in terms of itinerant quasiparticles with modified mass and energy. Recent studies of mixed-valence (MV) systems have led condensed matter physicists to explore the existence of exotic composite quasiparticles that are neutrally charged \cite{baskaran_majorana_2015, knolle_excitons_2017, erten_skyrme_2017, chowdhury_mixed-valence_2018, varma_majoranas_2020}. Intriguingly, these MV systems have the potential to produce Majorana fermions that do not require a superconducting state for their existence \cite{varma_majoranas_2020}.

These exotic effects arise from a peculiar interaction in MV systems that competes with the 4$f$$-$5$d$ hybridization in the standard Anderson model \cite{haldane_new_1977}. This interaction is described as a screening term, which is crucial for the fulfillment of the Friedel sum rule when two charge states coexist for a short time. However, the role of this interaction in MV systems remains unclear \cite{Haldane_ValIns_1977, Anderson_present_1984}, 
and was only considered in a handful of theoretical studies \cite{perakis_non-fermi-liquid_1993,varma_aspects_1994}.

More generally, there is a strong motivation to identify the various interactions between 4$f$, 5$d$, and perhaps the other states from anions in rare earth compounds \cite{takeshige_calculation_1985,Takeshige_Role_1991,allen_electronic_1986}. 
The MV compound investigated here, \TM\!\!, offers a unique opportunity to study the evolution of electronic interactions in response to changes in the lattice parameter using photoemission spectroscopy.
The conceptual view of \TM is as follows: the MV character sits on the Tm ion, which lies between Tm$^{2+}$ (4$f^{13}$) and Tm$^{3+} ($4$f^{12}$)  (see Supplementary Sec. S I). When the Se/Te composition is changed, the $fcc$ crystal structure is retained. Both anions (Se and Te) accept about two electrons from a Tm cation to form an (almost) closed shell; therefore, replacing Se with Te leads to lattice expansion without significantly affecting the periodicity of the Tm ions. This is largely analogous to changing the lattice constant by applying pressure. Transport and magnetic properties corroborate this analogy, as the behavior of \TM under hydrostatic pressure smoothly transitions to that of TmSe at ambient pressure \cite{derr_valence_2006, tsiok_general_2014,boppart_semiconductor-metal_1985}.

To understand what evolution can be expected when the Te/Se ratio is changed, it is instructive to look at the properties of the two end members, TmSe and TmTe. At one end, TmSe is the only binary MV system \cite{campagna_spontaneous_1974} that exhibits both antiferromagnetic and semimetallic properties \cite{moller_field-dependent_1977,clayman_effects_1977}.
At the other end, TmTe is almost an ionic insulator with a quadrupolar magnetic order \cite{matsumura_quadrupolar_1995, clementyev_neutron_1997}. Thus, a semimetal-insulator transition should occur between TmSe and TmTe, causing a drastic change in the 4$f$ bonding properties at ambient pressure. 

Two macroscopic effects can be observed in the transition that can be attributed to the difference in the 4$f$ bonding properties. First, there is a strong change in the compressibility, which is higher in the semimetallic phase. 
Second, there is a sudden change in the effective magnetic moments across the transition, indicating a significant difference in the 4$f$ occupation numbers. However, systematic spectroscopic studies of \TM as a function of $x$ that reveal the underlying electronic structure changes are still lacking. 

Here we report the detailed evolution of the electronic structure of \TM across the semimetal-insulator transition as obtained from core and valence electron emissions using synchrotron-based hard x-ray and extreme ultraviolet photoelectron spectroscopy (PES) \cite{kaindl_surface_1982,ufuktepe_resonant_1998,lebegue_electronic_2005}.
We firmly establish a bulk electronic transition at $x=0.29$ between semimetallic and insulating electronic states with distinct non-integer valences. In the semimetallic phase, our results reveal a conspicuous two-component 4$f$-peak structure indicating a so far omitted interaction of the 4$f$ electrons.

\textit{Experiment.}---Single-crystalline \TM samples were grown with Tm flux using the Bridgman method (Supplementary Sec. S II). The Te content ($x$) was mainly determined by energy-dispersive x-ray spectroscopy (EDXS).
Hard x-ray photoemission spectroscopy (HAXPES) and high-resolution extreme ultraviolet (XUV) photoemission spectroscopy were performed at ALS MERLIN, BESSY II 1$^{3}$, DIAMOND I09, PETRA III P22 \& P04, and SOLEIL TEMPO, respectively. Photoemission spectra were taken from cleaved (001) surfaces, typically within six hours after cleavage under ultrahigh vacuum, in a temperature range from 2~K to 110~K (See also Supplementary Sec. S III). 

%\section{Results}
%\subsection{Fig1}

\begin{figure}
\includegraphics[width=0.48\textwidth]{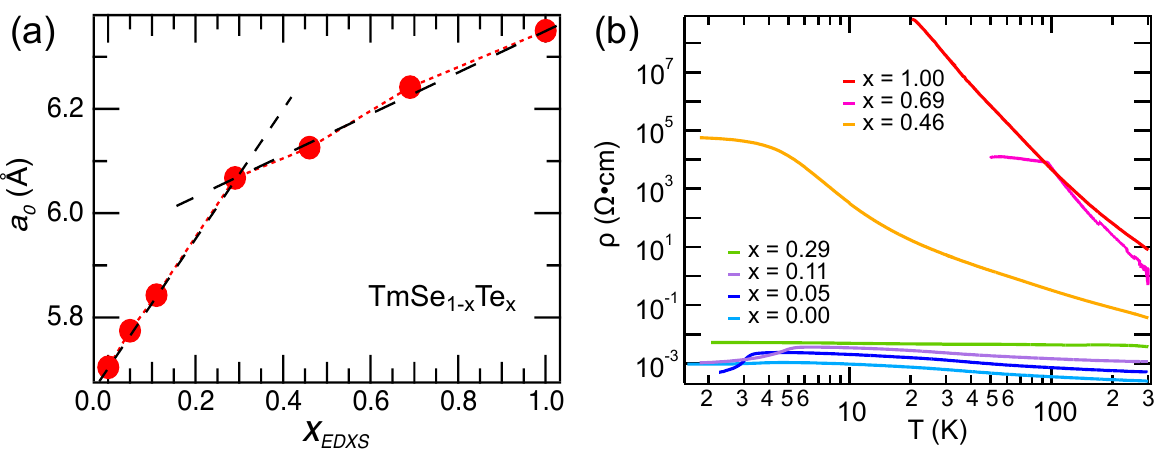}
	\caption{(a) Lattice parameter \Aa as a function of the Te content $x_\text{EDXS}$. Two fitted straight lines with different slopes (black dashed lines) intersect at $x_\text{EDXS} = 0.29$, indicating a phase transition. (b) Temperature- and Te concentration-dependent resistivity, showing a transition at around $x_{\rm EDXS} = 0.29$ from semimetallic behavior for $x_{\rm EDXS} \leq 0.11$ to semicondcuting behavior for $x_{\rm EDXS} \geq 0.46$.}
\label{FigResi}
\end{figure}

\textit{Electrical properties.}---Figure \ref{FigResi}(a) shows the relationship between the lattice constant \Aa and the composition ratio $x_\text{EDXS}$, which are measured after the crystal growth. In particular, $x_\text{EDXS}$ is hereafter denoted as $x$ (Supplementary Table I). As the two fitted lines show, there are two different regimes with different slopes. Since $x$ and \Aa can be considered as external pressure and volume, respectively, the different slopes indicate different compressibilities. The samples with lower $x$ values show higher compressibility, which is in agreement with the previous result reported in Ref. \cite{boppart_semiconductor-metal_1985}. Note that the intersection of the two lines is at $x = 0.29$.

The resistivity data of our samples in Fig.\,\ref{FigResi}(b) show the behavior known from the literature \cite{kaldis_phase_1979,kobler_intermediate_1981}.
For $x \leq 0.11$, the resistivity increases by about an order of magnitude with decreasing temperature $T$, but remains at a low level. This corresponds to the behavior of TmSe ($x = 0$), which is classified as a semimetal: The absolute electrical resistivity is low, while the optical reflectivity is high, and the material exhibits characteristic conduction band plasma effects at room temperature \cite{batlogg_dielectric_1981}. For $x > 0.29$, however, the resistivity increase with decreasing $T$ is much more pronounced, indicating semiconducting or insulating behavior. At the transition point $x = 0.29$, the temperature dependence of the resistivity is flat, representing an intermediate behavior. The data therefore establish a semimetal-insulator transition in \TM at an actual $x = 0.29$.

%\subsection{Fig2}
\begin{figure}%[t]
\centering
\includegraphics[width=0.48\textwidth]{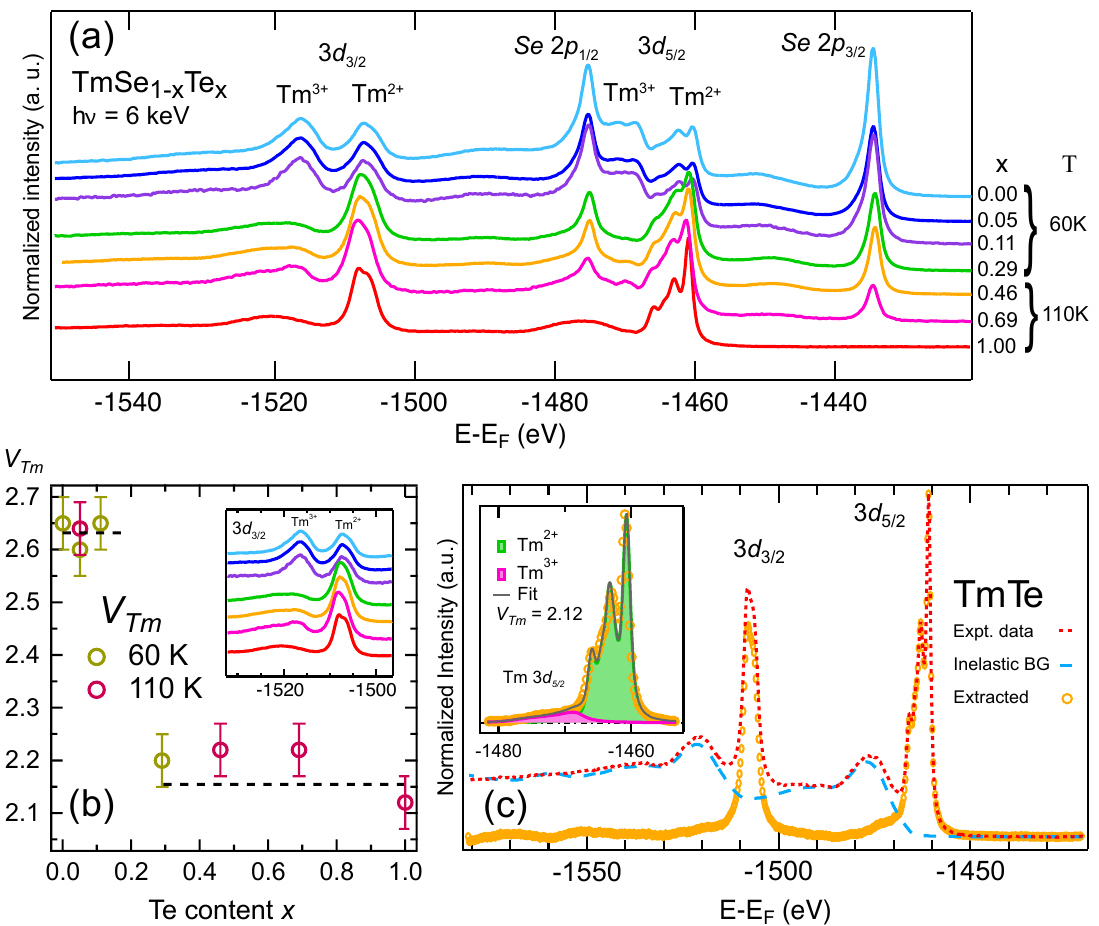}
	\caption{(a) Bulk-sensitive Se 2$p$ and Tm 3$d$ core-level spectra of \TM for different Te substitutions $x$. Due to the mixed-valence properties, all Tm core-level emissions consist of two ionization peaks. (b) Extracted Tm valence (\VTM\!\!), showing a strong deviation between two phases, namely around 2.65 and 2.15 (dashed lines), while remaining almost constant in each phase. For the semimetallic samples ($0 \leq x \leq 0.11$), the spectral weight of Tm$^{2+}$ and Tm$^{3+}$ emissions is comparable, while for the insulating samples (0.29 $< x \leq$ 1.0), that of Tm$^{2+}$ is dominant, as shown in the inset of (b) with 3$d_{3/2}$. (c) \VTM of TmTe: By deconvoluting the loss function from the Te 3$p$ spectrum as explained in Supplementary Sec.~\ref{sec:VTMestimation}, the intrinsic Tm 3$d$ spectrum (orange circles) is extracted. Inset of (c): With the intrinsic spectrum, the Tm$^{2+}$ and Tm$^{3+}$ contributions are identified (green and pink areas, respectively), giving a non-integer \VTM = 2.12 even in the insulating TmTe.}
	\label{FigCL}
\end{figure}

\textit{Mixed valence and valence transition.}---
In rare-earths systems, the electronic properties at the surface can substantially deviate from those in the bulk \cite{reinert_photoemission_1998,lutz_valence_2016,utsumi_bulk_2017}. We use HAXPES to probe the bulk valence properties. The insulating samples with $x \geq 0.46$ were measured at a higher temperature to avoid sample charging. Figure \ref{FigCL}(a) shows representative core-level emissions of Tm 3$d$ and Se 2$p$ with their spin-orbit partners (Supplementary Sec. S III). 
Both 3$d_{5/2}$ and 3$d_{3/2}$ peaks consist of two ionization peaks of Tm$^{2+}$ and Tm$^{3+}$ due to the MV nature.

The Tm valence \VTM is calculated from the average spectral weight of Tm$^{2+}$ and Tm$^{3+}$. The two Se 2$p$ peaks and the inelastic background must be removed to estimate the Tm$^{3+}$ spectral weight, since the latter is strongly reduced in the samples with $x \geq 0.29$, as shown in the inset of Fig. \ref{FigCL}(b). Figure \ref{FigCL}(b) shows the \VTM values obtained after subtracting these contributions (Supplementary Sec. S IV). The \VTM values differ between semimetallic and insulating samples, while they remain almost constant between the samples of the same phase. Such $x$-dependent \VTM behavior is consistent with magnetic susceptibility measurements \cite{kobler_intermediate_1981, boppart_semiconductor-metal_1985}, which do not show any integer \VTM in the Tm monochalcogenide. The \VTM values for the semimetallic and insulating samples are around 2.65 and 2.15 (indicated by the dashed lines), respectively, which can be converted to \nf $=$ 12.35, and \nf $=$ 12.85, respectively. We do not find a clear temperature dependence in \VTM\!\!, as exemplified by the two data points for $x = 0.05$ in Fig. \ref{FigCL}(b), unlike the case of the Sm valence in SmB$_6$ \cite{mizumaki_temperature_2009,lutz_valence_2016,utsumi_bulk_2017}. 
Often \VTM in the insulating phase has been considered to be divalent \cite{launois_x-ray_1980, jarrige_valence_2005}, but our estimation clearly shows a non-integer value. 
As shown in detail in Supplementary Sec. S IV, we can obtain an estimate of \VTM for the insulating TmTe [Fig.~\ref{FigCL}(c)] more easily because the loss function can be unambiguously determined using Te $3p$ core-level spectra. As a result, the inelastic background (cyan dashed line) can be removed and the intrinsic Tm $3d$ spectrum can be extracted (orange circles). From the intrinsic spectrum [inset of Fig.~\ref{FigCL}(c)], \VTM is evaluated using multiplet ligand field theory calculations \cite{haverkort_quanty_2016}. Our fitting results clearly yield a non-integer \VTM = 2.12 $\pm$ 0.05 that is consistent with bulk-sensitive magnetic susceptibility and spectroscopic measurements \cite{kobler_intermediate_1981,kinoshita_resonant_1998,launois_x-ray_1980, brewer_intermediate_1985}. Since our bulk-sensitive data were obtained on freshly cleaved surfaces, the non-integer \VTM cannot be attributed to contamination \cite{kaindl_surface_1982}.

Based on the fact that the conduction 5$d$ states are shifted above the Fermi level \EF to become insulating \cite{antonov_electronic_2001}, 
the non-integer valence indicates that the hybridization between 4$f$ and non-5$d$ states also exists in the insulating phase. We speculate that the Te 5$p$ states should be involved in the covalent bonding formation. Thus, it implies that the $f-p$ hybridization plays a role in the insulating phase.
%\subsection{Fig3}
\begin{figure}
%	\centering% \ref{fig:fig1}
\includegraphics[width=0.48\textwidth]{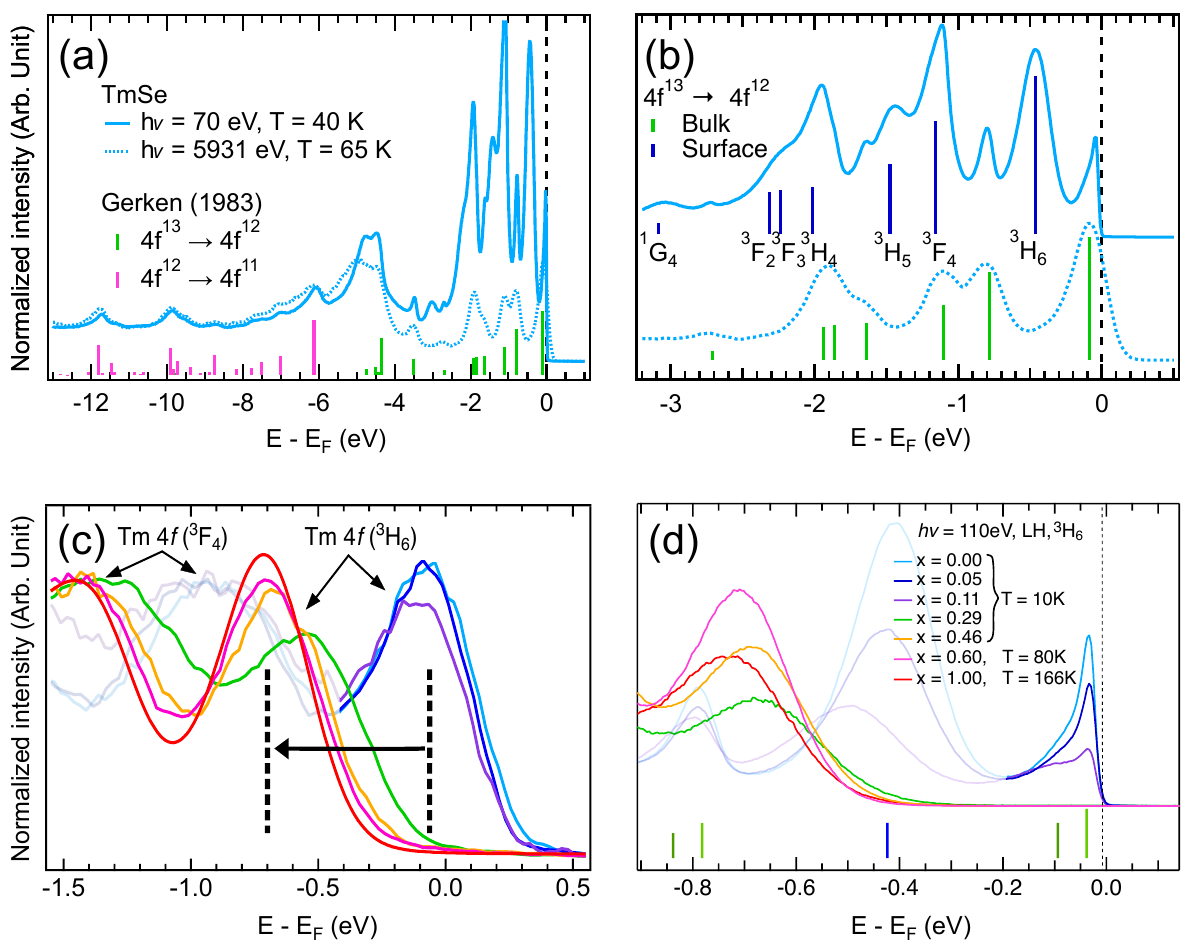}
	\caption{(a) Overall 4$f$ spectra of TmSe recorded at XUV and hard x-ray photon energies. Due to the mixed-valence nature, both 4$f^{12}$ (pink bars) and 4$f^{11}$ (green bars) final states appear as multiplet lines. (b) Enlargement of the 4$f^{12}$ final states in (a) near the Fermi level \EF\!\!. The surface contributions (blue bars) in the \hv $= 70$~eV data are stronger than the bulk peak contributions (green bars). Except for these surface peaks, both spectra show identical peak structures. 
(c) Hard x-ray Tm 4$f$ spectra were obtained at the same temperatures as in Fig. \ref{FigCL}. In semimetallic samples, the \HS peaks are positioned near \EF\!\!, while in insulating samples, they appear at lower energies. The maximum peak energy separation between TmSe and TmTe is approximately 0.6 eV (indicated by black dashed lines). Spectral weight in the gap region reaches a maximum at the transition point $x =$ 0.29. (d) XUV Tm 4$f$ spectra exhibit a similar trend to the HAXPES results in (c). The spectral weight in the gap region and the energy separation of the \HS peaks match those in (c).
 }
	\label{FigDOS}
\end{figure}

\textit{4f multiplet structure and spectral gap.}---We take a closer look at the 4$f$ emissions near \EF with both XUV and hard x-ray excitations. Figure~\ref{FigDOS}(a) shows the complete 4$f$ multiplet structure of TmSe. Due to the 4$f^{13}$ and 4$f^{12}$ contributions in the ground-state wave function, both 4$f^{12}$ and 4$f^{11}$ final states are accessible after removing one electron in the photoemission process. The green and pink bars represent the theoretically calculated atomic multiplet lines for the corresponding transitions \cite{gerken_calculated_1983}. The spectrum taken with XUV light (\hv = 70~eV) shows a more complicated peak structure than that at \hv = 6~keV because of additional surface 4$f$ peaks (Supplementary Sec. S V). However, we find that both the XUV and hard x-ray 4$f$ spectra are consistent, except for the presence of these surface peaks. Figure~\ref{FigDOS}(b) shows the enlarged 4$f$ spectra of the $f^{13} \rightarrow f^{12}$ transition. The surface peaks (blue bars) are dominant in the XUV spectra (cyan straight line) and are absent in the HAXPES data (cyan dashed line). The surface peaks are located at \EB $\approx -0.4$~eV on the freshly cleaved surface and shift to lower \EB with time [Supplementary Fig.~\ref{FigSuBu}(b)]. 

The first two bulk 4$f$ peaks, \HS and \Ff\!\!, in \TM [Figs.~\ref{FigDOS}(c) and \ref{FigDOS}(d)] do not overlap with the surface peaks, making it easier to analyze the bulk electronic states. Since the peak position of \HS determines the gap opening, the peaks below the \HS peaks of semimetallic samples are faintly depicted to emphasize the \HS peaks of the insulating samples (Supplementary Sec. S III). 
 In Fig.~\ref{FigDOS}(c), medium-resolution HAXPES spectra illustrate the evolution of the 4$f$ states across the semimetal-insulator transition. For the semimetallic samples ($0 \leq x \leq 0.11$), the \HS appears at \EB $\approx -50$~meV, while for the insulating samples ($0.46 \leq x \leq 1$), the \HS peaks are below $-0.5$~eV. Between TmSe ($x = 0$) and TmTe ($x = 1$), the energy difference in \HS becomes maximal ($\approx$ 0.6~eV), which can be considered as a fully opened gap. Furthermore, the \HS peak becomes broader up to $x = 0.29$ and narrower for larger $x > 0.29$. Thus, for $x = 0.29$, spectral weight still largely fills the gap region, indicating that this concentration is indeed close to the semimetal-insulator transition. It can be seen that the shapes of the \HS peaks are different among the three semimetallic samples. Therefore, we now investigate their behavior with higher energy-resolution using XUV photoemission spectroscopy. 

Figure \ref{FigDOS}(d) shows high-resolution 4$f$ spectra taken at \hv $= 110$~eV near \EF as a function of $x$ (Supplementary Sec. S III). 
The spectral features are mostly consistent with those of HAXPES shown in Fig. \ref{FigDOS}(c): The energy difference in \HS between semimetallic and insulating phases is about 0.6~eV, and the larger spectral weight in the gap region is present at $x = 0.29$. Therefore, we attribute the XUV \HS peaks to bulk states with higher energy resolution. The results show that the \HS peak in TmTe is located at about \EB $= -0.7$~eV; thus, the charge gap is larger than 0.3~eV \cite{suryanarayanan_optical_1975,boppart_semiconductor-metal_1985}. In addition, asymmetric shapes and intensity variation of \HS are observed among semimetallic samples.

%\subsection{Fig4}
\begin{figure}
%	\centering% \ref{fig:fig1}
\includegraphics[width=0.5\textwidth]{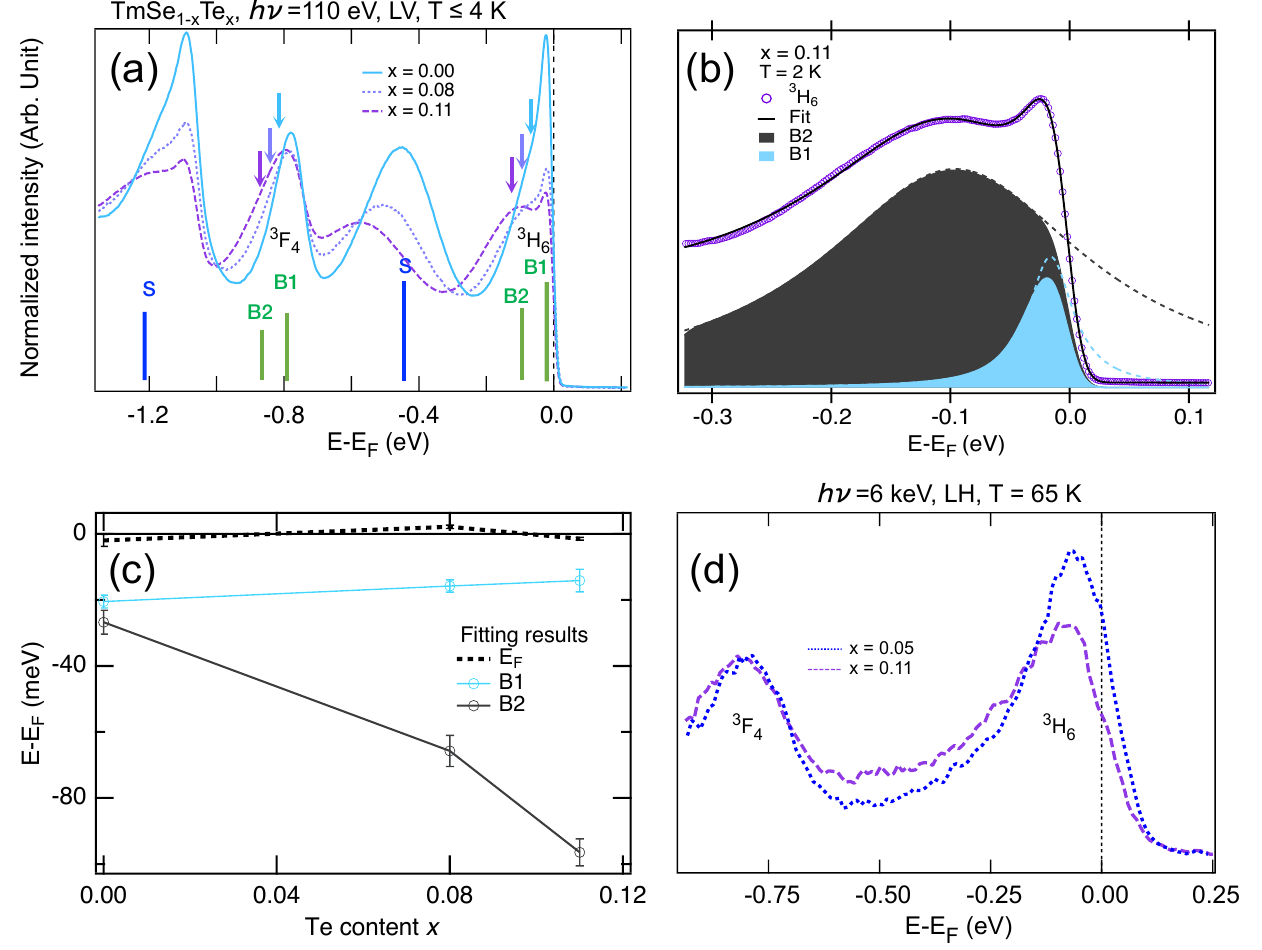}
	\caption{(a) Low-temperature 4$f$ spectra measured with linear vertical polarization to highlight the asymmetric peak shape of \HS and \Ff. With increasing $x$, both peaks become more asymmetric as indicated by the downward arrows at \HS and \Ff\!\!.
    (b) Line decomposition of  \HS\!\!. The asymmetric shape is due to the two-peak structure of each bulk multiplet peak. For example, the \HS peak for $x =$ 0.11 consists of a sharp peak (B1) and a broad peak (B2). As $x$ increases, the spectral weight of the B1 at \EF decreases relative to that of the B2, colored in cyan and black, respectively. (c) Energy difference between B1 and B2, increasing with $x$. The energy position of B1 remains almost the same, but B2 shifts to lower energies, as indicated by the arrows in (a), up to \EB $= -96$~meV. (d) High-resolution HAXPES spectra of \HS\!\! showing the increasing asymmetric line shape with increasing $x$, thereby supporting the two-peak structure in the bulk electronic structure.}
	\label{Fig2peak}
\end{figure}

\textit{4f fine structure in the semimetallic phase.}--- Low-temperature measurements ($T = 2$~K) were performed to study the asymmetric shapes at reduced thermal broadening. In addition, a different light polarization was used to reduce the intensity of the surface multiplet peaks, as shown in Fig.~\ref{Fig2peak}(a) (Supplementary Sec. S III). 
We also found that both the \HS and \Ff emissions become more asymmetric with increasing $x$, as indicated by the downward arrows. As a result, each peak consists of two peaks, B1 and B2, as indicated by the green bars.

Figure \ref{Fig2peak}(b) shows that the \HS emission for $x =0.11$ consists of a sharp peak (\EB $\approx -14$~meV) and a broad peak (\EB $\leq -96$~meV) whose splitting is too large to be attributed to crystal-field splitting in Tm monochalcogenides \cite{furrer_comparison_1981, matsumura_enhancement_2002}. Note that as $x$ increases, the sharp peak B1 is suppressed, while the broad peak B2 becomes relatively more intense and shifts to lower energies, as indicated by the downward arrows [see also Fig. \ref{Fig2peak}(a)]. The peak positions of B1 and B2 as a function of $x$ are plotted in Fig. \ref{Fig2peak}(c). The energy difference between B1 and B2 increases from 6~meV to 82~meV for a 2\% lattice expansion from $x = 0$ to $x = 0.11$. 

If B2 is from a bulk state, the two-peak structure should also be observable in bulk-sensitive HAXPES. Therefore, as additional evidence, we have taken high-resolution HAXPES spectra of the \HS peaks for $x = 0.05$ and 0.11, as shown in Fig.~\ref{Fig2peak}(d) (Supplementary Sec. S III). These spectra show asymmetric peaks consistent with the broadening of \HS observed in Fig.~\ref{Fig2peak}(c). Interestingly, the two-peak structure has also been observed in the other MV system SmB$_6$ \cite{min_two_2015}.

%\section{Discussion}

\textit{Discussion.}---Our comprehensive photoemission experiments on \TM across the MV semimetal-to-insulator phase transition show good agreement with resistivity data, providing compelling evidence for the phase transition at $x=0.29$. More interestingly, we have found strong distinct 4$f$ photoemission signatures between the semimetallic and insulating phases. The two main experimental observations are (1) a non-integer \VTM that decreases significantly across the transition and (2) a two-peak structure of \HS and \Ff in the semimetallic phase.

First, the non-integer \VTM in the insulating TmTe is discussed. \VTM $= 2.12$ in the insulating states indicates that 0.12 electrons have been shared from the Tm 4$f$ state. It indicates that some bonding between 4$f$ and other states still exists in the insulating phase.
 When the 5$d$ states are occupied, the system may enter a metallic phase. Thus, we can rule out the existence of the 5$d$ states, and one can naturally speculate that non-5$d$ states are involved in bonding with the 4$f$ states. A strong candidate is the anion $p$-band, which spreads 1.7 eV below \EF\!\!, consistent with covalent bonding causing non-integer \VTM\!\!.

Second, we expect a clear change in the bonding character as \VTM decreases significantly from semimetallic to insulating phases. The strong reduction in \VTM across the transition indicates that 0.5 electrons are mostly transferred from the 5$d$6$s$ states to the 4$f$ states. The consequence appears in the 4$f$ spectra with a large energy shift of 0.6~eV. It indicates strong changes in the 4$f$$-$5$d$ hybridization due to the absence of 5$d$6$s$ states below \EF\!\!.

Finally, we focus on the constant \VTM (\nf\!\!) and the two-peak structure among the semimetallic samples ($0 \leq x \leq 0.11$). In particular, we have found that the relative intensities of the two peaks vary with $x$. Since \nf remains unchanged (constant \VTM\!\!), these results indicate that the sharp 4$f$ states (B1) seem to transform into the broader 4$f$ states (B2) with increasing $x$ (lattice expansion). 
This suggests that the bonding properties of the 4$f$ states in \TM evolve with increasing $x$.

We also performed DMFT+DFT calculations, but the two-peak structure was not reproduced (Supplementary Sec. S VI). Although more in-depth studies are needed to understand the evolution from B1 to B2 states in MV systems, our novel experimental results have revealed an additional 4$f$ state that is absent in standard Anderson model studies \cite{denlinger_temperature_2013}. This finding provides evidence for the need for an additional interaction term, going beyond the standard Anderson model. For example, the 4$f$-5$d$ screening term \cite{Haldane_ValIns_1977, varma_majoranas_2020} can provide insight into the interpretation of the 4$f$ nature. In addition, the evolution of the coupling between 4$f$ electrons and phonons, leading to the formation of a polaronic state, could play a role in heavy fermion systems \cite{Martin_interpolative_2008}. While various interactions could be proposed, our results indicate that the desired interaction should induce an energy separation between B1 and B2 of up to 76 meV with a 2\% increase in the lattice parameter.

\textit{Conclusion.}---
We have investigated the nature of the localized 4$f$ states in \TM for a wide range of compositions. Our results highlight tunable interactions among the 4$f$ states as $x$ varies across the semimetal-insulator phase transition. These interactions across the transition extend beyond the hybridization with 5$d$6$s$ states, and include other valence orbitals. We have demonstrated this by resolving large variations in the non-integer \VTM\!\!. Moreover, we observe two distinct 4$f$ states in the low-energy semimetallic electronic structure. Notably, the population of these two states varies with $x$, even in the absence of changes in the 4$f$ occupancy. The energy splitting between these two states significantly depends on $x$, exhibiting a substantial 76 meV shift when the lattice parameter is expanded by 2\%. These observations reveal a rich and complex correlated electron phase in \TM\!\!, well beyond the predictions of the standard Anderson model that has served as a paradigm for the description of heavy-fermion compounds. The highly tunable character of the 4$f$ states in \TM may induce distinct physical phenomena, potentially promoting the formation of exotic quasiparticles. However, rigorous theoretical studies will be required to elucidate the underlying nature of the novel 4$f$ low-energy phenomenology we observe here in the metallic phase of \TM\!\!.

%\section{Acknowledgments}
\begin{acknowledgments}
\indent C.H.M. and L.D. would like to thank J. W. Allen and H.-D. Kim for their helpful discussions. C.H.M. would also like to thank K. Kissner, T. R. F. Peixoto, C. Fornari, J. Buck and W.-S. Kyung for their kind assistance. This work was supported by the German Research Foundation (DFG) in the framework of the Würzburg–Dresden Cluster of Excellence on Complexity and Topology in Quantum Matter–ct.qmat (EXC 2147, Project-ID 390858490), Project-ID 258499086–SFB 1170 (projects A01, C06) and BMBF Project-ID 05K19WW2. This research used resources of the Advanced Light Source supported by a DOE Office of Science User Facility under Contract No. DE-AC02-05CH11231. We acknowledge Diamond Light Source for access to beamline I09 (Proposals No. SI22630), which contributed to the results presented here. We acknowledge DESY (Hamburg, Germany), a member of the Helmholtz Association HGF, for the provision of experimental facilities. Parts of this research were carried out at PETRA III and we would like to thank the staff for their assistance.
Funding for the HAXPES instrument at P22 by the Federal Ministry of Education and Research (BMBF) under framework program ErUM is gratefully acknowledged.                                                                                                                                                                                                                          
We also thank the BMBF for funding the photoemission spectroscopy instrument at beamline P04 (contracts 05KS7FK2, 05K10FK1, 05K12FK1, and 05K13FK1 with Kiel University; 05KS7WW1 and 05K10WW2 with Würzburg University).
The research leading to the presented results was supported by the project CALIPSOplus under Grant Agreement 730872 from the EU Framework Program for Research and Innovation HORIZON 2020. Y.S.K. was supported by grants from the National Research Foundation (NRF) of Korea (NRF) funded by the Korean government (MSIP; Grant No.NRF-2019R1F1A1040989). C.-J.K. was supported by NRF (Grant No. 2022R1C1C1008200) and the KISTI Supercomputing Center (Project No. KSC-2022-CRE-0438). N.W was supported by the Deutsche Forschungsgemeinschaft (DFG; German Research Foundation) via CRC 925, Project ID 170620586(Project B2)
\end{acknowledgments}

\bibliographystyle{apsrev4-1}
%
%\bibliography{Tm}

\end{document}

%\title{Coherent many-body resonance state forming indirect band gaps  in mixed-valence \TM}
%\title{Two exotic many-body resonance peaks and conduction band deformation in mixed-valence \TM}
%\title{Near-gap spectra and Tm valence of mixed-valence \TM}
\title{Supplementary materials: Anomalous 4$f$ fine structure in \TM across the metal-insulator transition}
\maketitle 

%\section{Supplementary Material}
\beginsupplement

 \section{Overview of Thulium monochalcogenides}
\label{sec:overview}

\TM exhibits two prominent features: (1) an antiferromagnetic mixed-valent (MV) ground state \cite{moller_field-dependent_1977}, and (2) a metal-insulator transition (MIT) that encompasses a bosonic exciton-condensation phase during the transition \cite{kaldis_phase_1979,kobler_intermediate_1981,bronold_possibility_2006,wachter_exciton_2018}. The ground state of most \TM is characterized by trivalent (\VTM = 3) and divalent (\VTM = 2) configurations, both uniquely magnetic ($J$ $\neq$ 0) \cite{campagna_spontaneous_1974, nath_photoemission_2003}. \VTM can be converted to the number of 4$f$ electrons (\nf\!\!), thereby estimating the number of other outermost shell electrons (\nc\!\!). For Tm$^{3+}$ (Tm$^{2+}$), the occupation of the 4$f$ states is \nf = 12 (\nf = 13), while that of the conduction state is \nc = 1 (\nc = 0) with the formal configuration described as $f^{12}$$c^1$ ($f^{13}$$c^0$). By Hund's rule, the former and the latter have the $J$ = 6 and $J$ = 7/2 symmetry for the ground state, respectively. 
 
 Anion-replacement of Se by Te induces the lattice expansion together with the increase of the radius of Tm cations \cite{launois_x-ray_1980,mignot_single-crystal_2000}. This indicates the decrease of \VTM\!\!, namely toward the divalence \cite{kobler_intermediate_1981}. Semimetallic TmSe shows \VTM = 2.65~$\pm$~0.05 (\nf = 12.35~$\pm$~0.05), which indicates smaller \Aa = 5.705\;\AA~at room temperature~\cite{bianconi_x-ray_1981,kinoshita_resonant_1998}, whereas insulating TmTe shows \VTM = 2.05~$\pm$~0.05 (\nf = 12.95 $\pm$~0.05), which reflects greater \Aa = 6.35~\AA~\cite{suryanarayanan_optical_1975,kinoshita_resonant_1998}. Consequently, during the MIT from TmSe to TmTe, a substantial charge redistribution will happen from the conduction electron states \nc to the localized 4$f$ states \nf\!\!. It is not intuitive how such an insulating state can show the intermediate valence. Even in the most bulk-sensitive $L_3$-edge fluorescence-yield x-ray absorption result, Tm$^{3+}$ contributions do appear as a shoulder \cite{jarrige_valence_2005}. 
 
%In the synthesis phase diagram of \TM\!\!, a miscibility gap exists between TmSe and TmTe, making it challenging to obtain high-quality single crystals. The corresponding lattice parameters fall roughly between 5.85~\AA\; and 6.07~\AA. These parameters can only be accessed by applying hydrostatic pressure to samples with \Aa $>$ 6.07~\AA \cite{wachter_exciton_2018}. 
 
 Unlike the other MV compounds, long-range magnetic orderings occur in both semimetallic and insulating phases. Type-I antiferromagnetic (AFM) ordering appears in TmSe at \TN $=$ 3.2 K \cite{moller_field-dependent_1977} while antiferroquadrupolar and type-II antiferromagnetic orderings occur in TmTe at $T_{Q}$ $=$ 1.8 K and \TN $=$ 0.23 K, respectively \cite{matsumura_low_1998}. Due to the presence of both spin and valence fluctuations at low temperatures, scattering studies on TmSe and TmTe have revealed distinctive features below \TN\!\!, which have not been observed in other MV Ce, Sm, and Yb compounds and a metallic Ce system with a long-range magnetic ordering \cite{loewenhaupt_spin_1979,mignot_neutron_1995,nagao_studies_2010}. The major difference of \TM is that the total number of electrons per unit cell is odd (\nt = 13), whereas for other MV materials this number is even \cite{dumm_mobility_2005}. Moreover, the crystal field splittings are not considered in this study since they are smaller than 10 meV, which even in transport measurements is barely detectable \cite{furrer_comparison_1981, matsumura_low_1998}.

 %%%%%%%%%%%%%%%%%%%%%%%%%%%%%%%%%%%%%%%%%%%%%%%%
%%%%
\section{Single crystal synthesis and Te/Se ratio characterization} 
\label{sec:transport}

\begin{figure}
\includegraphics[width=0.48\textwidth]{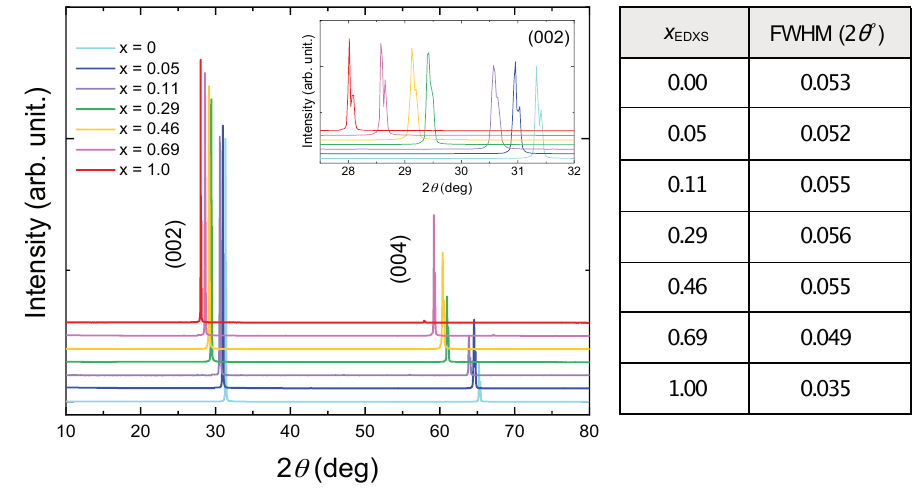}
	\caption{(Color online) X-ray diffraction patterns for the \TM single crystals.
	}
\label{FigXRD}
\end{figure}

Single crystals of \TM were grown by Bridgman method. The raw materials of Tm pieces (Alfa aesar 99.9\%), Se grains (Kojundo chemical, 99.999\%), and Te lumps (Alfa aesar 99.9995\%) were weighed in a molar ratio of Tm:Se$_{1-x}$Te$_x$ = 0.6 : 0.4. This method with self-flux Tm enables single crystal growth of \TM\!\!, which has a very high melting point of about 2500 $^\circ$C, at low temperatures. It has been known that the single crystal growth method at a low temperature has the advantage of reducing the number of crystal defects that otherwise may occur more frequently at a high temperature. The weighed materials were placed in a clean tungsten crucible baked at 2200 $^\circ$C. The crucible and the lid were then welded with an arc welder to prevent the volatility of the materials, especially Se and Te. The welded tungsten crucible was mounted in a high vacuum electric furnace consisting of a tungsten meshed heater with temperature stability of $\pm$ 0.1$^\circ$C. The single crystals were grown at 2000 $^\circ$C and the typical size was from 1$\times$1$\times$0.5 to 3$\times$3$\times$0.5 mm$^3$.

The x-ray diffraction (XRD) patterns of \TM single crystals were obtained using a PANalytical Empyrean diffractometer with Cu K$\alpha$ radiation ($\lambda_{K-\alpha_1}$=1.54060 and $\lambda_{K-\alpha_2}$=1.54443 \AA). As shown in Fig.\,\ref{FigXRD} (Left), only (00$l$) diffraction pattern was clearly observed, and as the amount of Te substitution increased, the peak positions shifted to a low angle due to an increase in the lattice constant. The lattice constant \Aa calculated by (00$l$) peaks is listed in Fig.\,\ref{FigXRD} (right), and the lattice constant increases as the amount of Te increases. The parent TmSe single crystal grown by our single crystal method, which prevents the volatility of Se, can exclude the sample inhomogeneity that may occur due to the Te substitution and Se defects. This is supported by the fact that the temperature dependence of the electrical resistivity of a few single crystals obtained from the same ingot was almost identical. As demonstrated in the inset to Fig. \,\ref{FigXRD} (Right), the homogeneous TmSe single crystal shows a very narrow (002) peak with FWHM = 0.0526 $^\circ$. 
Since the substitution by Te in TmSe may lead to inhomogeneity, FWHM of the XRD (002) peak for the substitution system \TM was evaluated and is listed in Fig.\,\ref{FigXRD} (Right). The FWHM for each $x$ is also very small and approximately equals that of TmSe, indicating that Te-substitution did not result in crystal inhomogeneity. Another evidence for the crystal homogeneity is that the temperature dependence of the electrical resistivity of a few single crystals (Fig. \ref{FigResi}) extracted from the same ingot is approximately the same. Therefore, the \TM single crystals used in this study did not cause inhomogeneity due to Te substitution. At a few single-crystal near the MIT, $i.e$, at $x_{\rm{EDXS}}$ = 0.11 and 0.29, we have sometimes found some pieces with slightly different Te concentrations compared to the other pieces synthesized at the same batch, but all their spectral features were fitting the general trends between $x_{\rm{EDXS}}$ = 0.05 and 0.46. Thus, all of them fit the general trend upon $x$. 

The composition ratio $x_{\rm{EDXS}}$ of each element in a single crystal of \TM was determined by Energy-Dispersive X-ray Spectroscopy (EDXS) and scarcely by Inductively Coupled Plasma (ICP) marked with $^{*}$ as shown in Table \ref{tax}.
The table shows the results of the compositional analyses of the grown \TM crystals, revealing significant deviations between the nominal values $x_\text{nominal}$, corresponding to the pregrowth mixing ratio, and the actual values $x_\text{EDXS}$ measured by EDXS. 
Such a discrepancy is common in crystal growth for multicomponent crystals \cite{Dudy_Yb_2013}.
Our $x_\text{nominal}$ values are close to the $x$ values reported in Ref. \cite{boppart_semiconductor-metal_1985} when comparing values for similar lattice constants \Aa\!\!.

\begin{table}[h!]
\begin{center}
\begin{tabular}{ |c||c|c| c| c| c |c |c| } 
 \hline
 $x_{\rm nominal}$ & 0.0 & 0.125 & 0.25 & 0.4 & 0.5 & 0.66 & 1.0 \\ 
 \hline
 $x_{\rm EDXS}$ & 0.0 & 0.05 & 0.11 & 0.29 & 0.46 & 0.69$^{*}$ & 1.0 \\ 
 \hline
 $a_{\rm 0}$ (\AA) & 5.705 & 5.775 & 5.843 & 6.068 & 6.126 & 6.242 & 6.366\\
 \hline
\end{tabular}
\end{center}
\caption{Nominal and measured Te contents $x_\text{EDXS}$ and lattice constants $a_0$ of \TM\!\!.}
\label{tax}
\end{table}

\Aa can be tuned quite continuously with varying $x$ except for the range of 5.85~\AA~$<$~\Aa~$<$~6.05~\AA~\cite{kobler_intermediate_1981,batlogg_valence_1981}. This exceptional range (miscibility gap) is near MIT, which might be an unstable phase region for crystal growth. Such isostructural first-order transition has been studied in Sm monochalcogenides as well \cite{deen_structural_2005}.  At $x$ = 0.29 (\Aa = 6.068 \AA). The reported activation gap values were 180 meV for $x$ $\sim$ 0.29 (\Aa = 6.07 \AA) \cite{batlogg_valence_1981}, and 110 meV for $x \sim$ 0.46 (\Aa = 6.14 \AA) \cite{boppart_semiconductor-metal_1985}.

%%%%%%%%%%%%%%%%%%%%%%%%%%%%%%%%%%%%%%%%%%%%%%%%%%
\section{Experimental and Normalization}
\label{sec:experimental}

Presumably, due to a strong charging effect, obtaining photoemission spectra of those insulating samples at $T <$ 70 K was impossible. For this reason, the data for these insulating samples were collected at 110 K. Additionally, the intensity of the incident light was varied, which allowed us to exclude any detectable charging at this higher temperature. All XUV 4$f$-electron spectra were taken in normal emission and angle-integrated over more than 40$^{\circ}$ along the analyzer slit direction, while HAXPES data were taken with the transmission lens mode of the electron analyzers. The 4$f$ multiplet lines result from the transition from the ground state to the one-electron removal final state. Each multiplet line represents a specific photohole state among different final state configurations, resulting in different net Coulomb interactions \cite{gerken_calculated_1983}. The closer the state is to \EF\!\!, the less energy is required to create it. 

Hard x-ray photoelectron spectroscopy (HAXPES) were performed with \hv $=$ 6~keV at the beamline P22 of PETRA III using a Phoibos 225HV electron analyzer \cite{schlueter_new_2019} with the total energy resolution ($\Delta E$) of 340~meV (Figs. \ref{FigCL}, \ref{FigDOS} (c) and \ref{FigTmTe}) and $\Delta E$ = 270~meV (Fig.\ref{Fig2peak} (d)). After cooling the cryostat with LHe, the base pressure was below 5.0$\times$10$^{-10}$ mbar. Theoretically, the kinetic energy of 3$d$ photoelectrons produced with \hv $=$ 6~keV has an inelastic mean free path of 6 nm. The core-level spectra in Fig. \ref{FigCL} are normalized to the total spectral weight of Tm 3$d_{3/2}$ after removing the Shirley background. The 4$f$ spectra in Fig. \ref{FigDOS} (c) and Fig.\ref{Fig2peak} (d)  are normalized with respect to the intensity of \Ff, and the \Ff peaks in the semimetallic phases are faintly displayed to empahsize the \HS peaks in the insulating states.

Additional measurements (see Fig.\;\ref{FigDOS} (a)) have also been performed at the beamline I09 of DIMOND (Didcot) equipped with the VG Scienta EW4000 analyzer using energy resolution of $\Delta E$ = 250~meV at \hv = 5931~eV. The base pressure was better than 5.0$\times$10$^{-10}$ mbar.

The high-energy-resolution photoemission spectroscopy (HR-PES) was carried out at the beamline MERLIN of ALS (Berkeley) with the energy resolution $\Delta E \leq$ 30 meV at \hv = 70, and 110~eV using the VG Scienta R8000 analyzer (Fig.\;\ref{FigDOS} (a-b, d), and Fig.\;\ref{FigSuBu}(b-c)). The base pressure was lower than 7.0$\times$10$^{-11}$ mbar. The 4$f$ spectra of Fig.\;\ref{FigDOS} (d) are normalized to match the intensities around \EB $= -3$~eV and $-3.5$~eV for metallic and insulating samples, respectively. Moreover, the surface \HS and bulk \Ff peaks are rendered faint to focus on the \HS peaks.  

The high-resolution spectra at $T \approx 2K$ shown in Fig.\;\ref{Fig2peak} (a-b) were achieved using the one-cube ARPES setup, where a Scienta-omicron DA30 analyzer installed at the UE112-PGM2b beamline of the BESSY-II. The total energy resolution was $\Delta E \leq$ 28 meV at \hv = 110~eV. The base pressure was lower than 1.0$\times$10$^{-10}$ mbar. The 4$f$ spectra in Fig.\;\ref{Fig2peak} (a) are normalized with respect to the intensity at \EB $\approx -3$~eV.

The PES spectra for $x$ = 0.29 and 0.46 in Fig.~\ref{FigSuBu}(c) were measured at $T$ = 40 K at the beamline TEMPO of SOLEIL (Gif-sur-Yvette) using a MBS A1 analyzer with the total resolution $\Delta E$ $=$ 40 meV at \hv = 110 eV. The base pressure was better than 5.0$\times$10$^{-10}$ mbar. The 4$f$ spectra in Fig.~\ref{FigSuBu}(c)  are normalized to the intensity of \Ff peaks.

Soft x-ray core-level spectroscopy in Fig.~\ref{FigSuBu} (a) was performed at the BL P04 of PETRA III (Hamburg) using \hv = 1800 eV with $\Delta E$ $\leq$ 400 meV where the ASPEHRE III endstation is equipped with a Scienta-Omicron DA30 analyzer. The base pressure was better than 2.0$\times$10$^{-10}$ mbar.

%%%%%%%%%%%%%%%%%%%%%%%%%%%%%%%%%%%%%%%%%%%%%%%%%%

\section{\VTM estimation} 
\label{sec:VTMestimation}

\subsection{valence of TmTe}
\begin{figure}
\centering
\includegraphics[width=0.48\textwidth]{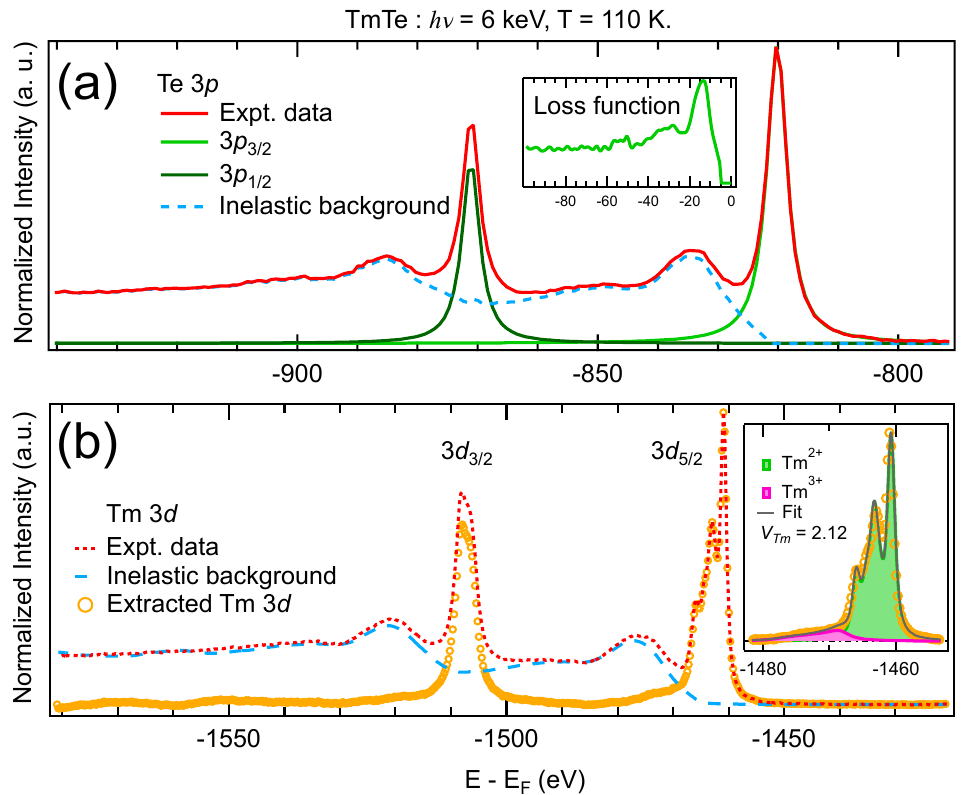}
\caption{HAXPES core-level study of TmTe (a) Simple Te 3$p$ core-level spectrum (red) is utilized to extract the full line shape of the energy loss function of photoelectrons (inset), which can generate the inelastic background shape (cyan dashed line). (b) By deconvoluting the loss function out of the experimental spectrum, we extract the intrinsic Tm 3$d$ spectrum (orange circles). (b, inset) With the intrinsic one, the Tm$^{3+}$ and Tm$^{2+}$ contributions are identified (pink and green areas, respectively), which gives a non-integer \VTM = 2.12 even in the insulating states.
}
\label{FigTmTe}
\end{figure}

We demonstrate our analysis procedures to obtain \VTM\!\! for insulating states. The Quanty program was used for the multiplet structure of the Tm core-level \cite{haverkort_quanty_2016} and deconvolute the loss function to remove the inelastic background (BG). The loss function describes various extrinsic energy losses of photoelectrons due to the inelastic scatterings while escaping from the sample. The inelastic BG is the result of a convolution of intrinsic core-level spectral functions and a loss function. 

The simple peak structure of Te $3p$ is used to deduce the loss function Fig.\;\ref{FigTmTe} (a, inset), which enables to generate the inelastic BG (cyan-dashed line). By deconvoluting the loss function out of the experimental Tm 3$d$ core-level spectrum, the inelastic BG is filtered out whereby the intrinsic $3d$ spectrum is extracted (orange circles in Fig.\;\ref{FigTmTe} (b)). From the intrinsic spectrum, \VTM are evaluated using the multiplet ligand field theory calculations (inset of Fig.\;\ref{FigTmTe} (b)) \cite{haverkort_quanty_2016}. Our fitting results clearly show the non-integer \VTM (= 2.12), which is consistent with a few bulk-sensitive magnetic susceptibility and spectroscopic results \cite{kobler_intermediate_1981,kinoshita_resonant_1998,launois_x-ray_1980, brewer_intermediate_1985}. Such intermediate valence indicates an interaction between 4$f$ and non-5$d$6$s$ states also exists in the insulating phase, which can be naturally connected to the covalent bonding formation between Tm and Te ions. Since our bulk-sensitive data were taken on freshly cleaved surfaces, it would be very challenging to attribute the non-integer valencies to the contamination \cite{kaindl_surface_1982}.  

% till here is from the main text.
\subsection{\VTM estimation for the semimetallic samples}

The general idea relies on fitting the experimentally measured Tm 3d core level spectrum to the sum of 2+ and 3+ components, while the components themselves are obtained within atomic multiplet calculation:
\begin{multline}
\chi^2(\alpha_2, \alpha_3) = \int \left( I_\text{exp}(E) -  I_\text{mod}(E)\right)^2  dE \rightarrow \text{min}, \\ 
\text{with }
I_\text{mod}(E) = \alpha_2 I_\text{mod}^{2+}(E) + \alpha_3 I_\text{mod}^{3+}(E)
\label{simple_fit}
\end{multline}
The required valence is then given by
\begin{equation}
v = \frac{2 \alpha_2 +  3 \alpha_3}{ \alpha_2 +   \alpha_3}.
\end{equation} 

Since the 4$f$ electrons are well shielded, the strength of the crystal field (CF) for the rare earth elements can be disregarded altogether because of the experimental resolution and purpose of this calculation \cite{zirngiebl_crystal-field_1984, Loewenhaupt1985245, Alistair066502, nickerson_physical_1971, jiao2016, Alekseev1993, Antonov165209}.  For the same reason, we can safely neglect the CF effects in
the $d$-shell, since the  3$d$ core orbitals are even more localized than the  4$f$ valence levels. Therefore, to properly describe the  experimental line-shape of the XPS spectrum it suffices to account in the Hamiltonian for the spin--orbit interaction in the 3$d$- and 4$f$-shells and for the Coulomb repulsion between 3$d$ and 4$f$ electrons \cite{ballhausen1962ligand, cowan1981theory}:
\begin{multline}
\hat{H} = \hat{H}^{dd}_\text{Coul} + \hat{H}^{df}_\text{Coul} + \hat{H}^{ff}_\text{Coul} \\
+ \zeta_{3d} \,\boldsymbol{\hat{L}}_{3d} \cdot \boldsymbol{\hat{S}}_{3d}
+ \zeta_{4f} \,\boldsymbol{\hat{L}}_{4f} \cdot \boldsymbol{\hat{S}}_{4f}.
\end{multline}
% <- do not remove this comment
For example, the Coulomb term $\hat{H}^{ff}_\text{Coul}$ can be relatively simply expressed as
\begin{equation}
\hat{H}^{ff}_\text{Coul} = 
\frac{1}{2}
\sum\limits_{p, q, u, t}   
%\sum\limits_{\begin{smallmatrix} p, q, \\u, t \end{smallmatrix}} 
\sum\limits_{k=0}^{l} a_k(p,q,u,t) F_{ff}^{(2k)}
\hat{f}^\dag_p
\hat{f}^\dag_{q}
\hat{f}_{u}
\hat{f}_{t},
\end{equation} 
where the outer sum runs over the $7 \times 2 = 14$ $4f$ spin-orbitals,  $a_k$ are known constants 
determined by the angular part of the 4$f$-states, so that $\hat{H}^{ff}_\text{Coul}$
 is fully determined by the four Slater integrals, $F_{ff}^{(0)}$, $F_{ff}^{(2)}$, $F_{ff}^{(4)}$, $F_{ff}^{(6)}$,  
which in turn depend on the radial part of the one-particle 4$f$-states.  Similarly, the $f$-$d$ part depends on five other parameters, controlling so-called direct ($F_{df}^{(2)}$, $F_{df}^{(4)}$) and exchange ($G_{df}^{(1)}$, $G_{df}^{(3)}$, $G_{df}^{(5)}$) Coulomb interactions between $f$- and $d$-orbitals \cite{ballhausen1962ligand}.

Eventually, the angle-integrated Tm 3d XPS spectrum can then be approximated by 
\begin{multline}
I(E) 
\sim
%\frac{1}{Z}
\sum_{i}
\sum_{\alpha} e^\frac{-E_\alpha}{k_\text{B}T}
\sum_{\beta} \left| \left \langle
\Psi^{N-1}_\beta | \hat{d_i} | \Psi^{N}_\alpha 
\right \rangle \right|^{2} \\
 \times
 \frac{-1}{\phantom{-}\pi}
\text{Im}\left[\frac{1}{(E_{\beta}-E_{\alpha}-E) + i\frac{\Gamma(E)}{2}}\right].
\label{eq_final}
\end{multline}
Here, the middle sum takes care of the thermodynamic averaging for a finite temperature $T$, including a possible degeneracy of the initial state $|\Psi_\alpha\rangle$. The inner sum accounts for the transition probabilities between the initial $N$-particle state $|\Psi^N_\alpha\rangle$, and the final $N-1$-particle state $|\Psi^{N-1}_\beta\rangle$, whereas the corresponding lifetimes are given by the function $\Gamma(E)$, which in the current consideration is to be determined from fits to experimental data. The outermost sum runs over the 10 3d spin-orbitals, with $\hat{d_i}$ being the corresponding annihilation operator.

\begin{figure}
%	\centering
\includegraphics[width=\columnwidth]{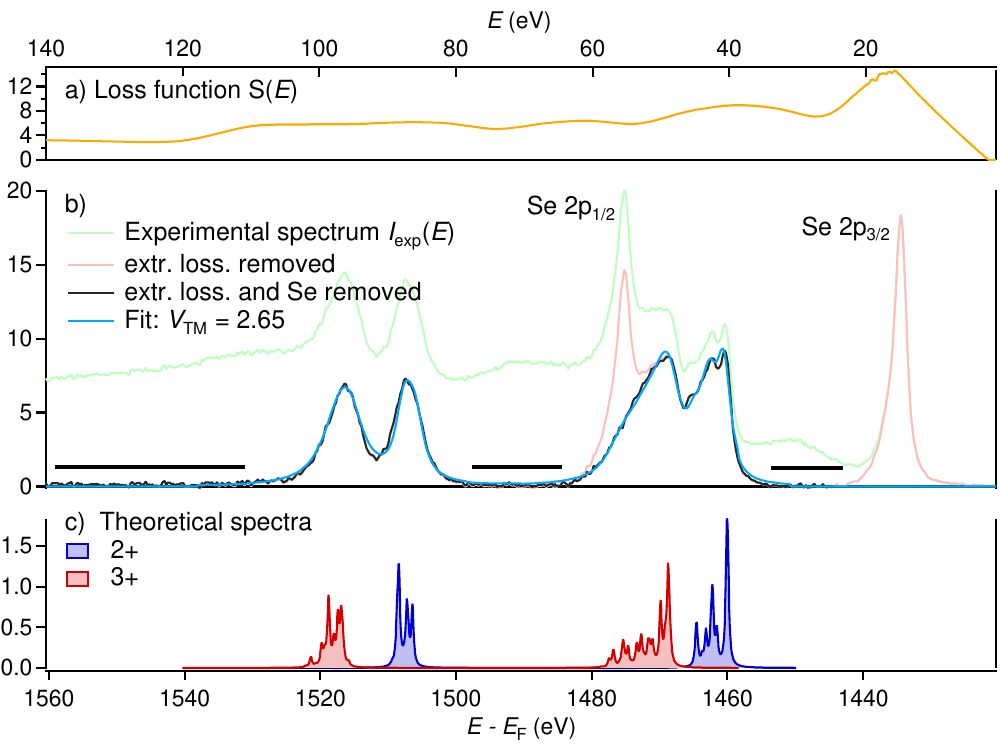}
	\caption{Extracting pure Tm 3$d$ core-level. (a) The loss function without the leading delta peak at $E=0$. (b) The process of deconvolution of the extrinsic losses and removal of Se 2$p$ peak. (c) Theoretical spectra for Tm$^{2+}$ and Tm$^{3+}$ were used to fit the deconvoluted data.}
	\label{TmSe3d}
\end{figure}

The actual calculations were performed using \textit{Quanty}  \cite{haverkort_quanty_2016, haverkort_multiplet_2012, Lu_efficient_2014, Haverkort_bands_2014}, a program developed by Prof. M. W. Haverkort, which offers a convenient and flexible way to define this quantum mechanical system in second quantization. The values for the spin--orbit couplings and the Slater integrals were obtained using the Cowan code \cite{cowan1981theory}. The usual scaling of the atomic Slater integrals down to 74--80\% of their atomic Hartree--Fock values was applied to account for intra-atomic screening effects. The resulting theoretical spectra with the realistic lifetime and experimental broadening excluded are shown in Fig. \ref{TmSe3d} (c). 

If it were not for the extrinsic losses and overlap with the Se $2p_{1/2}$ peak, one could have directly used the theoretical spectra in eq. \ref{simple_fit} to determine the valency. In practice one first  has to determine $I^\text{no loss}_\text{exp}(E)$ by deconvoluting the losses given by the loss function $S(E)$ and then subtract the Se $2p_{1/2}$ peak:
\begin{multline}
I^\text{lossy}_\text{exp}(E) =   \int\limits_0^{+\infty} I^\text{no loss}_\text{exp}(E-x)   S(x) dx, \text{where}\\
I^\text{no loss}_\text{exp}(E)  =   I_\text{exp}(E) +  I^{2p_{1/2}}_\text{exp}(E) +  I^{2p_{3/2}}_\text{exp}(E) 
\end{multline}

As shown from Fig. \ref{TmSe3d} (b), the removal of the  Se $2p_{1/2}$ peak from the deconvoluted spectrum $I^\text{no loss}_\text{exp}(E)$ can be achieved by subtraction of a shifted and scaled copy of Se $2p_{3/2}$ peak.

Since $S(E)$ is a smooth and slowly varying function of energy it can be parametrized using a spline with just a handful of nodes. Optimizing the node positions such that the deconvoluted spectrum becomes maximally close to zero within the three energy ranges marked with the horizontal bars in Fig. \ref{TmSe3d} (b), the $S(E)$ and the deconvoluted spectrum can be determined.

\section{Surface and bulk issues}
\label{sec:SurfBulk}

\begin{figure}
%	\centering
\includegraphics[width=0.48\textwidth]{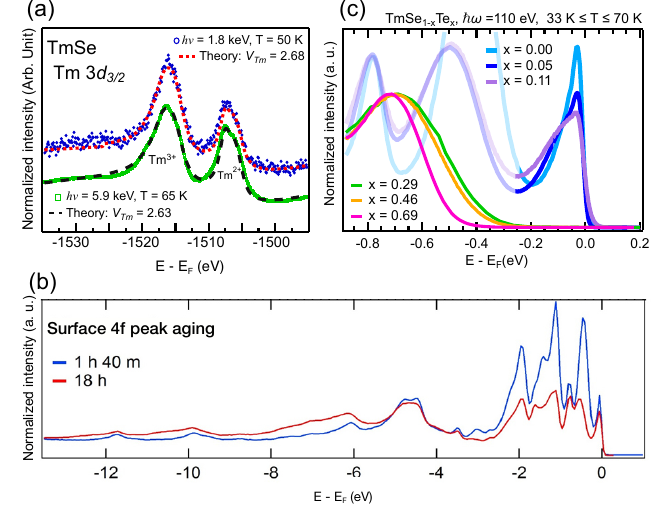}
	\caption{(Color online) (a) The Tm 3$d_{3/2}$ core-level spectra were obtained from freshly cleaved surfaces of TmSe with soft and hard x-rays. Estimated \VTM is almost identical disregarding the probing depth of our method, which indicates that both surface and bulk are mixed-valent states. (b) XUV-PES spectra show the aging effect on the 4$f$ spectra. Surface 4$f$ peaks are shown in (a) decrease in intensity with aging. (c) High-resolution PES spectra taken at \hv $=$110~eV reveal the same energy difference in the \HS peaks between semimetallic samples and insulating samples like the HAXPES data.}
	\label{FigSuBu}
\end{figure}

According to our analyses, both soft and hard x-ray spectra measured below 65~K give similar \VTM on freshly cleaved (001) surfaces (Fig.\;\ref{FigSuBu}(a)), but the former shows a slightly higher value. Even with bulk-sensitive HAXPES measurements, severe change in \VTM~starts to become apparent after 4 - 5~h after the cleaving. Hence, we have measured the Tm 3$d$ core-level spectra for each temperature by cleaving new single-crystalline samples and also using a low-energy resolution setup for quick data acquisition (Fig. \ref{FigCL}). We have also checked core-level spectra for several other single-crystalline pieces at the same $x$ and temperature with both XAS and PES to confirm that obtained \VTM values are reproduced for each $x$. Typical aging signature is the reduction of the surface 4$f$ multiplet peaks as shown in Fig.\;\ref{FigSuBu}(b); moreover, the Tm$^{3+}$ contribution increases with time. 

Figure \ref{FigSuBu}(c) shows 4$f$ spectra taken with \hv $=$ 110~eV near the \EF as a function of $x$, where surface \HS and bulk \Ff peaks are weakly rendered. The spectral features are therein mostly consistent with the HAXPES ones shown in Fig.\;\ref{FigCL} (c): The energy difference in \HS between semimetallic and insulating phases is about 0.6 eV, and $x =$ 0.29 shows the largest spectral weight in the gap region. With a further increase of $x$, the spectral weight in the gap region reduces. 

The main differences between XUV and hard X-ray data are found in the $x =$ 0.29. In the HAXPES, the \HS peak appears at \EB $= -$ 0.52 eV, whereas in the XUV spectra at $-$ 0.62 eV (Fig.\;\ref{FigSuBu} (c)). Depending on the cleaved surface the 4$f$ peak of $x =$ 0.29 appears at different energies in the range of $\pm$ 0.1 eV. However, the $x$ = 0.29 certainly shows the highest spectral weight in the gap region (-0.7 eV $<$ \EB $<$ 0.0 eV). As a result, one can conclude that a substantial decrease in the spectral weight happens inside the gap region upon $x$ among insulating states.    
\\
%\begin{figure}
%	\centering
%\includegraphics[width=0.3\textwidth]{FigS5.pdf}
%	\caption{Our core level analyses of Te 4$d$ and Se 3$d$ spectral weight (x-axis) obtained with \hv $=$ 128~eV have been proportional to the EDXS values (y-axis). Using this relationship, we estimate the "$x$" value of the sample assigned with $x > 0.11$ in Fig.\;\ref{FigDOS}(d), thereby resulting $x =$ 0.16.}
%	\label{FigTeSe}
%\end{figure} 

%As shown in Fig.\;\ref{FigSuBu}(c), some single-crystalline pieces of $x =$ 0.11 have different stoichiometry at least on the cleaved surface. To identify the $x$ value for those deviated from $x =$ 0.11, we have carried on the Te 4$d$ to Se 3$d$ core-level analyses for all \TM\!\!. By calculating the spectral weight ratio of Te 4$d$ to Se 3$d$ measured at \hv $=$ 128~eV, $x$ of those exceptional samples has been estimated (Fig.\;\ref{FigTeSe}). After obtaining the spectral ratios obtained from our core-level measurements (y-axis), we make a match with our EDXS values (x-axis). By fitting with a polynomial equation (blue line), the $x >$ 0.11 in the Fig.\;\ref{FigSuBu}(d) is identified to be $x$ = 0.16. Thus, our few pieces of single-crystalline sample synthesized under the same condition of $x =$ 0.11 have shown a slightly higher Te substitution rate but maintained in the semimetallic phase.  

%%%%%%%%%%%%%%%%%
\section{Computational details of DFT+DMFT}
\label{sec:band}

\begin{figure}
\includegraphics[width=0.47\textwidth]{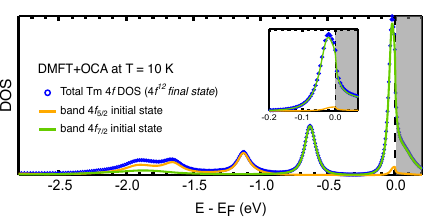}
\caption{Our DFT+DMFT (OCA impurity solver) results reproduce the bulk multiplet peaks for the 4$f^{12}$ final state at 10 K, but the two-peak structure is absent.}
\label{FigGGA}
\end{figure}

Our detailed investigation of the band structure of \TM will be published elsewhere. Here, we briefly mention our theoretical results based on the density functional theory plus dynamical mean-field theory (DFT+DMFT) scheme, which explains the electronic structure of strongly correlated systems well. First, in the DFT part, we have employed the all-electron full-potential linearized augmented plane wave (FLAPW) band method implemented in the WIEN2K package \cite{Blaha_wien2k_2019}.
A generalized gradient approximation (GGA) functional was chosen for exchange-correlational functional.
The spin-orbit coupling (SOC) is included as the second variational method.
Since antiferromagnetic ordering temperature $T_N$ is very low (3.5 K) and ARPES measurements were mostly done above $T_N$ or without sufficient energy resolution to probe the effect, we focus on a paramagnetic phase in this study. 
Based on the DFT with GGA+SOC scheme, we obtained a similar band structure of TmSe in Ref. \cite{jansen_local-density_1985}. Afterward, using the DFT results, we also performed DFT+DMFT calculations similar to Ref. \cite{kang_topological_2015} and analyzed the bulk 4$f^{12}$ final state peaks. To solve the impurity solver, we used the vertex corrected one-crossing approximation (OCA), in which the full atomic interaction matrix was taken into account \cite{cowan_1981}. The Coulomb interaction U = 6.0 eV, and
the Hund's coupling J = 1.0 eV were used for the DFT+DMFT calculations.
Overall 4$f^{12}$ final state peaks are nicely reproduced (Fig. \ref{FigGGA}), but the fine structure of B1 and B2 has been absent.

\bibliographystyle{apsrev4-1}
%\bibliography{Tm}
%